\documentclass[aps,superscriptaddress,amsmath,amssymb,preprintnumbers,showpacs]{revtex4}

\usepackage{graphicx}
\usepackage{mathrsfs}
\usepackage{amssymb}
\usepackage{bm}

\newcommand{\nslash}{\kern 0.2 em n\kern -0.50em /}
\newcommand{\kslash}{\kern 0.2 em k\kern -0.45em /}
\newcommand{\pslash}{\kern 0.2 em p\kern -0.50em /}
\newcommand{\Sslash}{\kern 0.2 em S\kern -0.50em /}
\newcommand{\Pslash}{\kern 0.2 em P\kern -0.50em /}
\newcommand{\Dslash}{\kern 0.2 em D\kern -0.65em /\kern 0.15em}
\newcommand{\slim}{\mskip 1.5mu}

\begin{document}

\title{The $\sin2\phi$ azimuthal asymmetry in single longitudinally polarized $\pi N$ Drell-Yan process}
\newcommand*{\SEU}{Department of Physics, Southeast University, Nanjing
211189, China}\affiliation{\SEU}
\newcommand*{\UTFSM}{Departamento de F\'\i sica, Universidad T\'ecnica
Federico Santa Mar\'\i a, and Centro Cient\'\i fico-Tecnol\'ogico de
Valpara\'\i so Casilla 110-V, Valpara\'\i so,
Chile}\affiliation{\UTFSM}
\newcommand*{\PKU}{School of Physics and State Key Laboratory of Nuclear Physics and
Technology, \\Peking University, Beijing 100871,
China}\affiliation{\PKU}
\newcommand*{\CHEP}{Center for High Energy
Physics, Peking University, Beijing 100871,
China}\affiliation{\CHEP}

\author{Zhun Lu}\affiliation{\SEU}\affiliation{\UTFSM}
\author{Bo-Qiang Ma}\email{mabq@pku.edu.cn}\affiliation{\PKU}\affiliation{\CHEP}
\author{Jun She}\affiliation{\PKU}

\begin{abstract}
We study the $\sin2\phi$ azimuthal asymmetry in the $\pi N$
Drell-Yan process, when the nucleon is longitudinally polarized. The
asymmetry is contributed by the combination of the Boer-Mulders
function and the longitudinal transversity distribution function. We
consider the Drell-Yan processes by $\pi^\pm$ beams colliding on the
proton and deuteron targets, respectively. We calculate the
$\sin2\phi$ azimuthal asymmetries in these processes using the
Boer-Mulders function and the longitudinal transversity from spectator
models. We show that the study on single polarized $\pi N$ Drell-Yan
processes can not only give the information on the new three-dimensional
parton distribution functions in momentum space, but also shed light
on the chiral-odd structure of the longitudinally polarized nucleon.
\end{abstract}

\maketitle

\section{introduction}

Transverse momentum dependent (TMD) distribution functions, or
alternatively named as three dimensional parton distribution
functions (3dPDFs) in momentum space, as an extension of the usual
Feynman distribution functions, enter the description of various
semi-inclusive reactions~\cite{bdr,Barone:2010ef}. The 3dPDFs, as
well as the three-dimensional fragmentation functions in momentum space,
encode a wealth of new information on the nucleon
structures~\cite{sivers,anselmino95,bhs02,collins02,belitsky,Boer:2003cm,
Ji:2006ub,Ji:2006vf,Ji:2006br,Bacchetta:2008xw} that cannot be
described merely by the leading-twist collinear picture. At leading
twist, there are eight 3dPDFs appearing in the decomposition of the
quark-quark correlation matrix of the
nucleon~\cite{Mulders:1995dh,Boer:1997nt,Bacchetta:2006tn}
\begin{align}
\Phi(x,p_T) &= \frac{1}{2}\, \biggl\{
f_1 \nslash_+
- {f_{1T}^\perp}\, \frac{\epsilon_T^{\rho \sigma} p_{T\rho}^{}\slim
  S_{T\sigma}^{}}{M} \, \nslash_+
+ \left(S_L\,g_{1L} - \frac{p_T \cdot S_T}{M}\,g_{1T}\right) \gamma_5\nslash_+
\nonumber \\[0.2em] & \quad \qquad
+h_{1T}\,\frac{\bigl[\Sslash_T, \nslash_+ \bigr]\gamma_5}{2}
+ \left(S_L\,h_{1L}^\perp - \frac{p_T \cdot S_T}{M}\,h_{1T}^\perp\right) \,\frac{\bigl[\pslash_T, \nslash_+ \bigr]\gamma_5}{2 M}
+i \, {h_1^\perp} \frac{ \bigl[\pslash_T, \nslash_+ \bigr]}{2M}
\biggr\}.
\label{eq:cor}
\end{align}
Here $n_+ = (0,1,\bm{0}_T)$ is a lightlike vector expressed in the
light-cone coordinates, in which an arbitrary four-vector $a$ is
written as $\{a^-,a^+, \boldsymbol{a}_T\}$, with $a^{\pm}=(a^0 \pm
a^3)/\sqrt{2}$ and $\boldsymbol{a}_T =(a^1,a^2)$. The distribution
functions on the right-hand sid (rhs) of (\ref{eq:cor}) depend on the longitudinal
momentum fraction $x$ and the square of the transverse momentum
$\boldsymbol{p}_T^2$, {\it i.e.}, the three-dimensional information
in momentum space.

Each of these eight 3dPDFs represents a special parton structure of
the nucleon. Five of them, the Sivers function $f_{1T}^\perp$, the
Boer-Mulders function $h_1^\perp$, the pretzelosity $h_{1T}^\perp$,
the distributions $g_{1T}$ and $h_{1L}^\perp$, vanish upon
integration in $\boldsymbol{p}_T$. The essential tools to explore
3dPDFs are the azimuthal asymmetries in various polarized or
unpolarized processes involving at least two hadrons, such as
semi-inclusive deeply inelastic scattering
(SIDIS)~\cite{Airapetian:2004tw,2009ti,Airapetian:2010ds,compass,compass06,Alekseev:2010rw,Mkrtchyan:2007sr,:2008rv,
emc87un, Breitweg:2000qh,compass2008un, compass2009un,hermes2009un}
and Drell-Yan~\cite{NA10,:2007mja,Zhu:2008sj} processes. The
experimental data have been applied to extract some of the 3dPDFs,
{\it i.e.}, the Sivers
function~\cite{anselmino05a,anselmino05b,efr05,cegmms,vy05} and the
Boer-Mulders
function~\cite{Zhang:2008nu,Lu:2009ip,Barone:2009hw,Barone:2010gk}.

Here we focus on the distribution $h_{1L}^\perp (x,p_T^2)$. It
describes the probability of finding a transversely polarized quark
inside a longitudinally polarized nucleon. So we could call it
longitudinal transversity (or shortly, longi-transversity or
heli-transversity). Therefore, this distribution manifests the
chiral-odd parton structure of a longitudinally polarized nucleon.
It has been shown~\cite{Mulders:1995dh,Bacchetta:2006tn} that
nonzero $h_{1L}^\perp (x,p_T^2)$ can yield a $\sin 2 \phi_h$
azimuthal asymmetry in the SIDIS process when the target nucleon is
longitudinally polarized. The model calculations of $h_{1L}^\perp
(x,p_T^2)$ have been given in
Refs.~\cite{Jakob:1997,Pasquini:2008,Bacchetta:2008af,Efremov:2009,Avakian:2010,Bacchetta:2010si,Zhu}.
The phenomenological studies of the $\sin2\phi_h$ asymmetry in the
longitudinally polarized SIDIS
process~\cite{Airapetian:1999tv,Airapetian:2002mf,Avakian:2010ae}
have been performed in \cite{Ma:2001,Ma:2002} and recently
in~\cite{Zhu,Boffi:2009sh}, showing that the asymmetry is around
several per cent.

In this paper, we consider the effect of $h_{1L}^\perp (x,p_T^2)$ in
the Drell-Yan process. As demonstrated in
Refs.~\cite{Boer1999,Arnold2009}, the combination of $h_{1L}^\perp
(x,p_T^2)$ and the Boer-Mulders function may lead to a $\sin 2\phi$
azimuthal asymmetry in the longitudinally polarized Drell-Yan
process, where $\phi$ is the azimuthal angle of the dilepton with
respect to the hadron plane. The phenomenological study of this kind
of asymmetry in Drell-Yan has not been presented in literature
yet. It is thus worth analyzing the $\sin 2\phi$ asymmetry in the
Drell-Yan process, which is the aim of this paper. The process we
consider is the $\pi N^\Rightarrow \rightarrow \ell^+\ell^- X$
process,
with the symbol ``$\Rightarrow$" denoting the longitudinal
polarization.
The transversely polarized and unpolarized $\pi N$ Drell-Yan processes have been studied in Refs.~\cite{Efremov2005,Bianconi2006,Lu:2006ew,Anselmino2009}.
The advantage of the $\pi N$ Drell-Yan process is that the
valence quarks participate the hard scattering, which might produce
a larger asymmetry than that in the $pp$ Drell-Yan process.
We point out that, except the combination of $h_{1L}^\perp (x,p_T^2)$
and the Boer-Mulders function, there is no competing mechanism for
the $\sin2\phi$ asymmetry in the single polarized Drell-Yan process,
even by considering higher-twist effect. Therefore the study of the
$\sin2\phi$ asymmetry provides a rather clean probe for the
distribution $h_{1L}^\perp (x,p_T^2)$.

\section{Single longitudinal-spin asymmetry in Drell-Yan process}
For a general Drell-Yan process with one of the incident hadrons longitudinally
polarized, i.e., $h_1 h_2^\Rightarrow\rightarrow \ell^+ \ell^- X$, the
single longitudinal-spin asymmetry may be defined as
\begin{eqnarray}
A_{UL}=\frac{d\sigma^{h_1 h_2^\Rightarrow\rightarrow \ell^+ \ell^- X} -
d\sigma^{h_1 h_2^\Leftarrow\rightarrow \ell^+ \ell^-
X}}{d\sigma^{h_1 h_2^\Rightarrow\rightarrow \ell^+ \ell^- X} +
d\sigma^{h_1 h_2^\Leftarrow\rightarrow \ell^+ \ell^- X}} \equiv
\frac{d\sigma^\Rightarrow - d\sigma^\Leftarrow}{d\sigma^\Rightarrow +
d\sigma^\Leftarrow}.
\end{eqnarray}
This definition is similar to that of the single transverse-spin
asymmetry. We treat the process in the parton model and only
consider the leading-order approximation via a single photon
transfer, {\it i.e.},
$q\bar{q}\rightarrow\gamma^*\rightarrow\ell^+\ell^-$. We denote the
momenta of the hadrons, the annihilating partons, and the produced
lepton pairs as $P_i$, $p_i$ and $k_i~(i=1,2)$, respectively. Then
the momentum transfer gives the invariant mass of the lepton pair
\begin{eqnarray}
q^2=(p_1+p_2)^2=(k_1+k_2)^2=M^2.
\end{eqnarray}
Now we work in the center of mass frame of two hadrons and
parameterize the four-momentum of the photon as
$q=(q_0,\bm{q}_T,q_L)$. At extremely high energies, if we assume
that the longitudinal component is dominant and neglect all the mass
effects and the transverse momenta, we can define the following
variables,
\begin{eqnarray}
&&x_1=\frac{q^2}{2P_1\cdot
q}\approx\frac{q_0+q_L}{\sqrt{s}},~~~~x_2=\frac{q^2}{2P_2\cdot
q}\approx\frac{q_0-q_L}{\sqrt{s}},\nonumber\\
&&\tau=\frac{M^2}{s},~~~~x_F=x_1-x_2\approx\frac{2q_L}{\sqrt{s}}.
\end{eqnarray}
Then we can build up the relation
\begin{eqnarray}
&&x_1=\frac{1}{2}\big(x_F+\sqrt{x_F^2+4\tau}\big),\nonumber\\
&&x_2=\frac{1}{2}\big(-x_F+\sqrt{x_F^2+4\tau}\big).\label{x}
\end{eqnarray}

The direction of the detected lepton pair can be described by the
solid angle ($\theta,\phi$), which is frame dependent. In the entire
paper, we will select the Collins-Soper frame~\cite{CS_frame}. In
the $h_1 h_2^\Rightarrow$ Drell-Yan process, if the transverse
momentum of the dilepton $\bm{q}_T$ is measured, we can apply the
TMD factorization~\cite{Collins1985,Ji2004,Collins:2004nx,Ji2005} to
write down the differential cross-section in the region $q_T^2\ll
M^2$ as~\cite{Boer1999,Arnold2009}
\begin{eqnarray}
\label{cross_section}
&&\frac{d\sigma}{d\Omega dx_1 dx_2
d^2\bm{q}_T}=\frac{\alpha^2}{3q^2}\bigg{\{}A(y)\mathcal{F}[\bar{f}_1f_1]\nonumber\\
&+&\left.S_{2L}B(y)\sin(2\phi)\times
\mathcal{F}\left[\frac{2(\hat{\bm{h}}\cdot\bm{p}_{1T})(\hat{\bm{h}}\cdot\bm{p}_{2T})
-\bm{p}_{1T}\cdot\bm{p}_{2T}}{M_1M_2}\bar{h}_1^\perp
h_{1L}^\perp\right]\right\}+\cdots, \label{eq:csul}
\end{eqnarray}
where $\cdots$ stands for the higher-twist contributions which will
not be considered in this paper.
 In the above expression we have used the notation $\mathcal{F}$ defined as
\begin{eqnarray}
\mathcal{F}[\omega \bar{f} g]&\equiv&\sum_{a,\bar{a}}\int
d\bm{p}_{1T}d\bm{p}_{2T}
\delta^2(\bm{p}_{1T}+\bm{p}_{2T}-\bm{q}_{T})\omega(\bm{p}_{1T},\bm{p}_{2T})\nonumber\\
&\times& \bar{f}^{\bar{a}}(x_1,\bm{p}_{1T})g^{a}(x_2,\bm{p}_{2T}),
\end{eqnarray}
and
\begin{eqnarray}
A(y) =  \frac{1}{2} - y + y^2  \stackrel{\mbox{cm}}{=}\frac{1}{4}( 1
+ \cos^2 \theta),~~~~~B(y) = y (1-y) \,
\stackrel{\mbox{cm}}{=}\frac{1}{4} \sin^2 \theta  ,
\end{eqnarray}
where $y = l^-/q^-$, with $l$ the momentum of the lepton, and
$\hat{\bm{h}} =\bm{q}_T/q_T$. At leading twist there are two
structure functions in the single longitudinally polarized Drell-Yan
process, as shown in (\ref{eq:csul}). The second structure function
shows that the combination of the two 3dPDFs, the Boer-Mulders
function and $h_{1L}^\perp (x,p_T^2)$, can lead to a $\sin2\phi$
azimuthal angle dependence of the dilepton. The size of this
azimuthal angle dependence can be obtained by defining the following
weighted asymmetry
\begin{eqnarray}
A^{\sin(2\phi)}_{UL}(x_1,x_2,y,q_T)&=&\frac{2\int_0^{2\pi}d\phi
\sin(2\phi)[d\sigma^\Rightarrow -
d\sigma^\Leftarrow]}{\int_0^{2\pi}d\phi [d\sigma^\Rightarrow +
d\sigma^\Leftarrow]} \nonumber\\
&=& {B(y)
\mathcal{F}\left[\left((2(\hat{\bm{h}}\cdot\bm{p}_{1T})(\hat{\bm{h}}\cdot\bm{p}_{2T})
-\bm{p}_{1T}\cdot\bm{p}_{2T})/M_1M_2\right)\bar{h}_1^\perp
h_{1L}^\perp\right]\over A(y)\mathcal{F}[\bar{f}_1f_1] }.
\end{eqnarray}
In this paper, we will investigate the $x_F,~M$ and $q_T$
dependences of this asymmetry, thus we need to change the variables
from $x_1$ and $x_2$ to $x_F$ and $M$ by using the relation
(\ref{x}). If the $\sin2\phi$ asymmetry can be measured by
experiment, it will provide a clear test on the chiral-odd structure
of the longitudinally polarized nucleon and will bring valuable
information of $h_{1L}^\perp (x,p_T^2)$. In Ref.~\cite{Zhu}, this
new 3dPDF, as well as another 3dPDF $g_{1T}$, was studied in the
SIDIS process. However, $g_{1T}$ can only be probed through double
spin asymmetry, thus it can not be studied in the $\pi N$ Drell-Yan
process.

\section{Numerical calculation}
In order to see the prospect of experimental measurements on the
single longitudinal-polarized Drell-Yan process, we calculate the
$\sin2\phi$ asymmetry of dilepton production by $\pi$ beams
colliding on the longitudinally polarized nucleons. The experiment
could be conducted at COMPASS of CERN in the near
future~\cite{Takekawa2010}, since there are already longitudinally
polarized proton and deuteron targets available; besides, the
utilization of the $\pi^-$ beam is quite promising at COMPASS.

We estimate the $\sin 2\phi$ asymmetry of the $\pi^\pm p^\Rightarrow$
and the $\pi^\pm d^\Rightarrow$ Drell-Yan processes at COMPASS. Although
the original proposal for the Drell-Yan process at COMPASS is to use
the $\pi^-$ beam, we consider the Drell-Yan process with $\pi^+$
beam as a supplement. Also we provide the calculations for both the
proton target and the deuteron target for comparison.

For the proton 3dPDFs, we will use the results obtained in a
light-cone quark spectator diquark model~\cite{Ma,Zhu} with the
relativistic Melosh-Wigner effect~\cite{MaOld} of quark transversal
motions taken into account. The 3dPDFs deduced from this model are
applicable in the hadronic scale. To compare with experimental
observables which are usually measured at rather high energies, it
is essential to evolve the parton distributions to the experimental
 scale from the model scale. However, here we calculate the azimuthal asymmetries
which are the ratios of different parton distributions, so the
effects of evolution are assumed to be small. In practice, this
model has been applied to obtain the helicity and transversity
distributions, which are reasonable to describe data related to
helicity distributions in a number of processes~\cite{Chen2005}, as
well as those related to the Collins asymmetry at
HERMES~\cite{Huang2007}.  This model is also successful in the
prediction of the dihadron production asymmetry at
COMPASS~\cite{She2008,compass_dihadron}. Therefore, it is worth trying to
apply the same model to the Drell-Yan kinematics at COMPASS.

For the deuteron 3dPDFs, we will adopt the simple assumption that a
deuteron nucleus consists of a free proton and a free neutron with
the same polarization. We have
\begin{eqnarray}
\label{TMD_D} f_a^D=f_a^p+f_a^n,
\end{eqnarray}
which holds not only for unpolarized distributions but also for
 polarized distributions. This may be different from the $^3$He
case which was discussed in Ref.~\cite{He3}, where, for the polarized
case, there should be effective polarization factors for different
nucleons because inside $^3$He the neutron and the two protons
have different polarizations. However, inside deuteron, the proton
and neutron have the same polarization, or, equivalently, we assume
that the effective polarization factor for each nucleon is 1. In
practice, we will also apply the isospin symmetry,
\begin{eqnarray}
f_u^D=f_d^D=f_u^p+f_u^n=f_u^p+f_d^p. \label{isospin}
\end{eqnarray}

Besides this, we need the Boer-Mulders function for
pions~\cite{Lu,Gamberg:2009uk}, and we will use the parametrization
in Ref.~\cite{Lu}, which was obtained in a quark spectator antiquark
model. The pion parton distributions we adopt were
demonstrated~\cite{Lu} to give a good description on the $\cos
2\phi$ asymmetries measured in the unpolarized $\pi N$ Drell-Yan
process~\cite{NA10}, where a large and increasing asymmetry was
observed in the $q_T$ region below 3 GeV; thus our model has been
checked to be reasonable in this region.

\begin{figure}
\includegraphics[width=\textwidth]{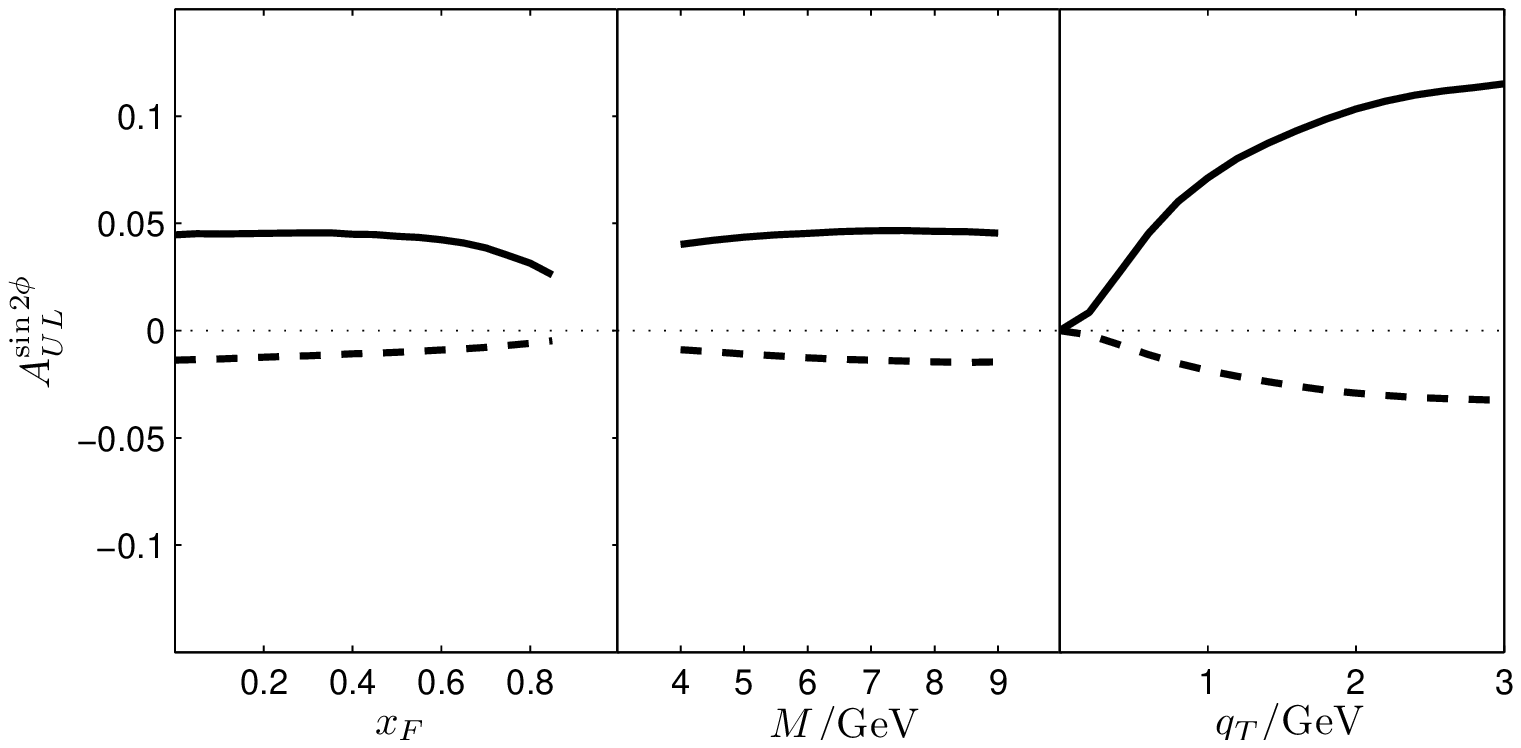}
\caption{The $\sin2\phi$ asymmetries for $\pi^{\pm}
p^\Rightarrow\rightarrow\mu^+\mu^-X$ processes at COMPASS. The solid and
dashed curves are the results for $\pi^-$ and $\pi^+$ beams,
respectively.} \label{prot}
\includegraphics[width=\textwidth]{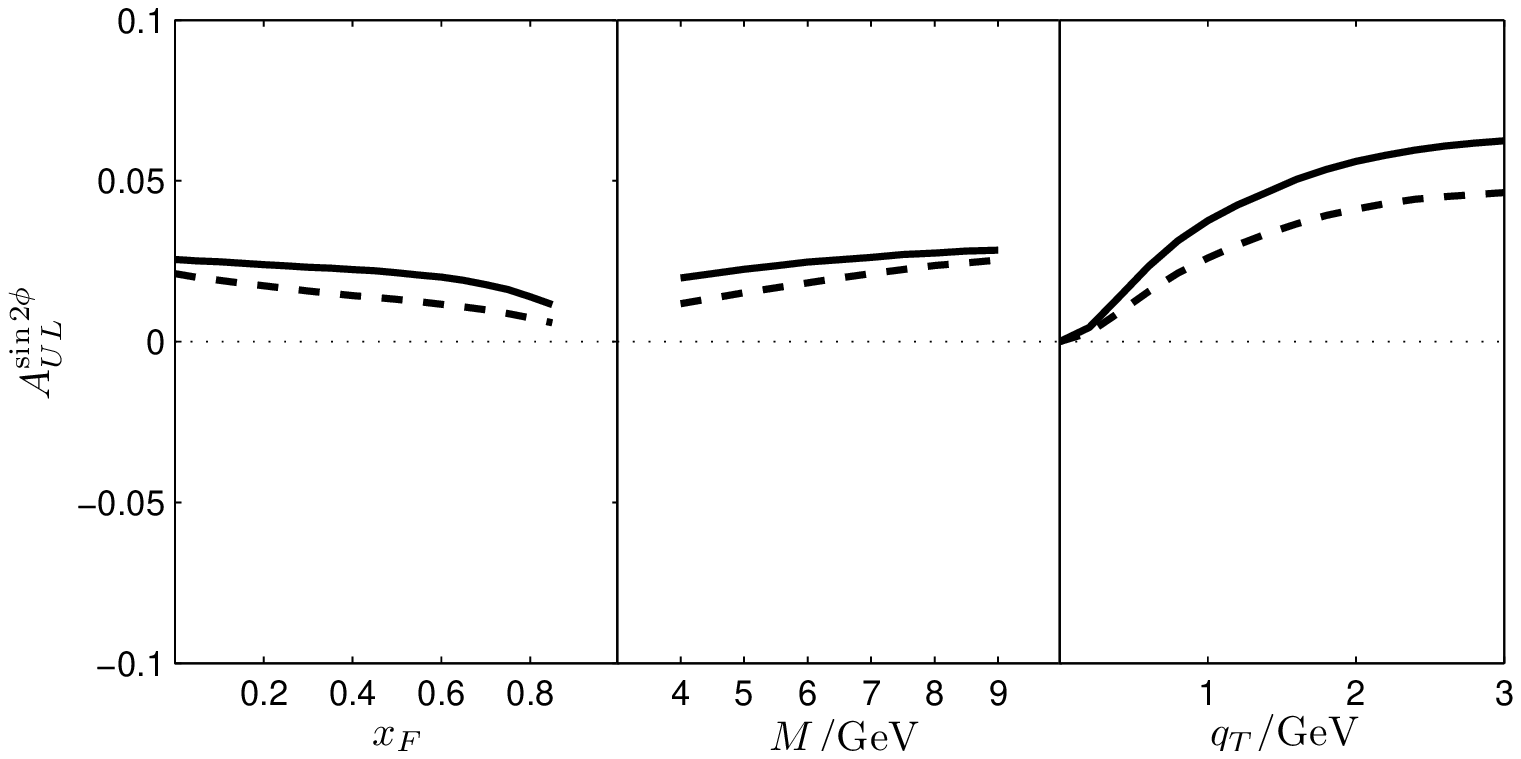}
\caption{Similar to Fig.~\ref{prot} but for the longitudinal-polarized deuteron target.} \label{deut}
\end{figure}

\begin{figure*}
\begin{center}
\scalebox{0.95}{\includegraphics[0pt,50pt][390pt,260pt]{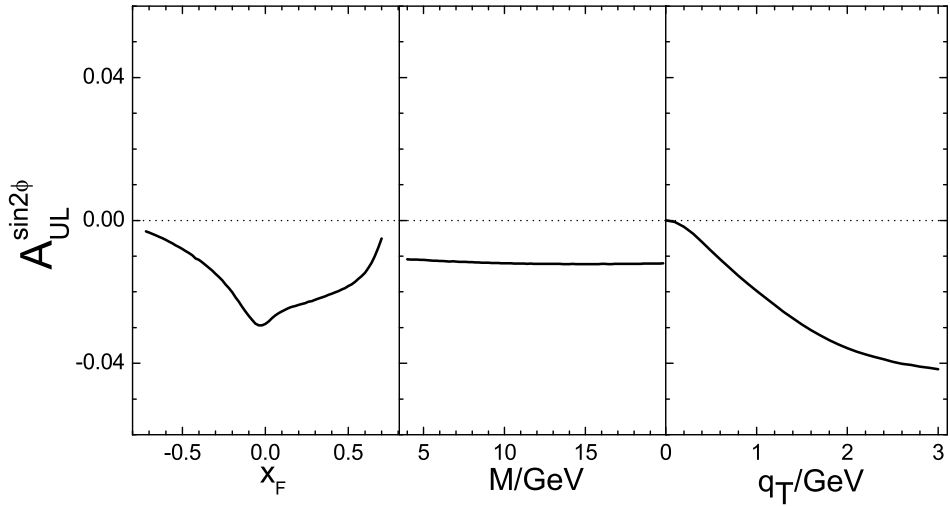}}
\caption{The $\sin2\phi$ asymmetries for $p
p^\Rightarrow\rightarrow l^+l^-X$ process at RHIC.} \label{rhiceps}
\end{center}
\end{figure*}

The COMPASS kinematics we adopt in the calculation
are~\cite{Takekawa2010}
\begin{eqnarray}
&&\sqrt{s}=18.9~\mathrm{GeV},~0.1<x_1<1,~0.05<x_2<0.5,\nonumber\\
&&4\leqslant M\leqslant8.5~\mathrm{GeV},~0\leqslant q_T \leqslant
4~\mathrm{GeV}~(\mathrm{if}~q_T~\mathrm{is~integrated}).\nonumber
\end{eqnarray}
We will give a detailed explanation for the integration ranges of the kinematical variables as
follows.
\begin{itemize}
\item For the $x_F$ dependence, we only give the prediction for forward
region $x_F>0$. Given a fixed $x_F^0$, the range for $M$ is
determined by Eq.~(\ref{x}) so that
$x_{1,2}^{\mathrm{min}}<x_{1,2}(x_F^0,M)<x_{1,2}^{\mathrm{max}}$.
\item For the $M$ dependence, given a fixed $M_0$, the range for $x_F$ is
determined by Eq.~(\ref{x}) so that
$x_{1,2}^{\mathrm{min}}<x_{1,2}(x_F,M_0)<x_{1,2}^{\mathrm{max}}$.
\item For the $q_T$ dependence, the range for $M$ is $4\leqslant
M\leqslant8.5~\mathrm{GeV}$ and the range for $x_F$ is determined
by Eq.~(\ref{x}) so that
$x_{1,2}^{\mathrm{min}}<x_{1,2}(x_F,M)<x_{1,2}^{\mathrm{max}}$.
\end{itemize}

We plot the $\sin2\phi$ asymmetries in
the $\pi p^\Rightarrow$ and $\pi D^\Rightarrow$ Drell-Yan processes at COMPASS in
Figs.~\ref{prot} and Fig.~\ref{deut}, respectively.
As the COMPASS experiments
will mainly probe the forward $x_F$ region, that is, the acceptance of COMPASS at negative $x_F$
is much lower than that at positive $x_F$, we only plot the asymmetries at the forward $x_F$ region.
It is interesting to see that the asymmetries are about several percent, especially for the $\pi^-$ beam.
Therefore, it is promising that the $\sin 2\phi$ asymmetries can be measured by experiments
with a rather good accuracy. This is
similar to what obtained in the SIDIS process~\cite{Zhu}, where sizable $\sin 2\phi_h$ asymmetry was
predicted.
So we do not need to apply the transverse momenta cutoff method to enhance the
asymmetry, as was applied in the calculation $\sin(3\phi-\phi_S)$ asymmetry in the transversely
polarized SIDIS~\cite{Shejun2009} and Drell-Yan processes~\cite{Lu:2011qp}. For the
deuteron target, because of the isospin symmetry, we obtain similar
results for the $\pi^+$ and $\pi^-$ beams. The sign of the $\pi d^\Rightarrow$
asymmetry is the same as that of the $\pi^- p^\Rightarrow$
asymmetry. This can be accounted for by the $u$ quark dominance in the
proton (See Eq.~(\ref{isospin}). The signs of the asymmetry in $\pi^+ p^\Rightarrow$ process and
that in $\pi^- p^\Rightarrow$ process are opposite, which is in contrast to the case in
 $\pi^{\pm} d^\Rightarrow$ process.

As a comparison, we calculate the $\sin 2\phi$ asymmetry for the $pp^\Rightarrow$ Drell-Yan
process at RHIC~\cite{Bunce:2000uv} with $\sqrt{s}=200~\textrm{GeV}$. In this case the pion beam is replaced by the proton beam.
The RHIC kinematics covers negative $x_F$ region, where the production of lepton pair is dominated by the sea antiquarks of the beam and valence quarks of the polarized proton. The kinematical cuts adopted in the calculations are
 \begin{eqnarray}
 4<M<20~\textrm{GeV},~~~ -2<{1\over 2}\ln\left({x_1\over x_2}\right)<-2,~~~ \textrm{and}~~~
 0<q_T<3~\textrm{GeV}.
\end{eqnarray}
In the calculation we need the Boer-Mulders functions of sea quarks
inside the proton. In order to make the calculation to be consistent
with the previous calculations for $\pi N$ Drell-Yan process, we
apply the baryon-meson fluctuation model results ~\cite{Lu:2007kj}
for $h_1^{\perp \bar{q}}$. In this model $h_1^{\perp \bar{q}}$ of
the proton is expressed as the convolution of the fluctuation
probability of $p\rightarrow$ baryon+$\pi$ meson and the
Boer-Mulders functions of the pion. In Fig.~\ref{rhiceps} we plot
the predicted asymmetries at RHIC as functions of $x_F$, $M$ and
$q_T$, respectively. The predicted asymmetries are of several
percent and are negative, similar to the case of $\pi^+
p^\Rightarrow$ process. We also present the asymmetry at the
negative $x_F$ region, which is sizable. The similarity of the
asymmetries in $\pi p^\Rightarrow$ and $p p^\Rightarrow$ processes
suggests that the measurements of $\sin2\phi$ asymmetry can also be
performed at RHIC.

\section{Conclusion}
We have presented the $\sin2\phi$ single spin asymmetries for the
$\pi^\pm N^\Rightarrow\rightarrow\mu^+\mu^-X$ processes at COMPASS.
The magnitude of the asymmetries is several percent, indicating that
they can be measured by experiments with a rather good accuracy. We
predict asymmetries for both proton and deuteron targets. Because of
the isospin symmetry, the $\pi^+$ and $\pi^-$ beams lead to the
similar
 magnitude of the asymmetry for deuteron target. The $\sin2\phi$
asymmetry in $\pi N^\Rightarrow$ Drell-Yan processes
 can not only provide the information on the Boer-Mulders
function for the pion but also shed light on the there-dimensional
parton distribution function $h_{1L}^\perp$, which tells us the
probability of finding a transversely polarized quark inside a
longitudinally polarized nucleon. This is why it was named as
longi-transversity or heli-transversity, as suggested in
Ref.~\cite{Zhu}. Our predictions have already demonstrated the
feasibility to probe this distribution at COMPASS; therefore, we
expect the experiment could perform corresponding measurement to
enrich our knowledge on the spin structure of the nucleon.

\section*{Acknowledgement}
This work is partially supported by National Natural Science
Foundation of China (Nos.~10905059, 11005018, 11021092, 10975003,
11035003) and by FONDECYT (Chile) under Project No.~11090085.

\end{document}